\documentclass{article}
 \usepackage[dvipdf]{epsfig}
  \usepackage{color}
\usepackage{graphicx}% Include figure files
\usepackage{color}
\usepackage{amssymb}

\def\z{{\zeta }}

\def\be{\begin{equation}}
\def\ee#1{\label{#1}\end{equation}}

 \def\bp{\mathbf{p} }
 \def\p{\textsf{p} }

\def\no{\nonumber}
\def\lb{\label}

\newcommand{\ben}{\begin{eqnarray}}
\newcommand{\een}{\end{eqnarray}}

\begin{document}

\title{Relativistic gas in a Schwarzschild metric}
\author{Gilberto M.  Kremer\footnote{kremer@fisica.ufpr.br}
\\Departamento de F\'{\i}sica, Universidade Federal do Paran\'a\\ Curitiba, Brazil}
\date{}
\maketitle

\begin{abstract}
 A relativistic gas in a Schwarzschild metric is studied within the framework of a relativistic Boltzmann equation in the presence of gravitational fields, where Marle's model for the collision operator of the Boltzmann equation is employed. The transport coefficients of bulk and shear viscosities and thermal conductivity are determined from the Chapman-Enskog method. It is shown that the transport coefficients depend on the gravitational potential. Expressions for the transport coefficients in the presence of weak gravitational fields in the non-relativistic (low temperatures) and ultra-relativistic (high temperatures) limiting  cases are given. Apart from the temperature gradient the heat flux has two relativistic terms. The first one, proposed by Eckart, is due to the inertia of energy and represents an isothermal heat flux when matter is accelerated. The other, suggested by Tolman, is proportional to the gravitational potential gradient and indicates that -- in the absence of an acceleration field -- a state of equilibrium of a relativistic gas in a gravitational field  can be  attained only if the temperature gradient is counterbalanced by a gravitational potential gradient.
\end{abstract}

%\pacs{51.20.+d, 04.20.-q, 05.20.Dd,  05.60.-k}

\section{Introduction}

The research on relativistic gases by using the Boltzmann equation is an old subject in the literature. We can state that the statistical description of a relativistic gas began with the works of
J\"uttner in 1911 and 1926, when he succeeded to derive the equilibrium distribution functions for a relativistic gas that obeys the Maxwell-Boltzmann \cite{J1} and the Fermi-Dirac and Bose-Einstein statistics \cite{J2}. The covariant formulation of the Boltzmann equation came later and was first proposed by  Lichnerowicz and Marrot \cite{LM} in 1940. The establishment of the non-equilibrium distribution function and of the transport coefficients of a relativistic gas by using the Chapman-Enskog method was done by Israel \cite{I} and Kelly \cite{K} in the sixties of the last century.    These pioneers works  were followed by several research papers in the literature concerning the study of gases in non-equilibrium states  by using the Boltzmann equation in special relativity. However, there exist only few papers in the literature concerning the  research of relativistic gases in the presence of gravitational fields within the framework of the Boltzmann equation. The first works are due to Chernikov  \cite{Ch1,Ch2}, who analyzed the equilibrium distribution functions in some specific metric tensors and  Bernstein \cite{B}, who determined the bulk viscosity for a relativistic gas in a Friedmann-Robertson-Walker metric.

Some years ago the influence of gravity on the heat transport in a rarefied gas  was analyzed in the works \cite{And1,And2} on the basis of a non-relativistic Boltzmann equation, where it was shown that a gravitational field parallel (anti-parallel) to the temperature gradient increases (decreases) the heat transport. These results are related with the one derived by Tolman \cite{To1,To2} from a general relativity theory, who showed  that a state of equilibrium of a relativistic gas in a gravitational field can be attained only if the temperature gradient is counterbalanced by a gravitational potential gradient.

Recently a relativistic gas in a gravitational field by using the Boltzmann equation was studied in the work \cite{AL}, where the dependence of the heat flux on the gravity -- the so-called Tolman's law -- was obtained. In this work no metric tensor was assigned -- although the components of the metric tensor appear explicitly in the equilibrium distribution function  -- and the dependence on the gravitational field was connected with the Christoffel symbol written in the Newtonian approximation.

The aim of the present work is to analyze a relativistic gas in a Schwarzschild metric within the framework of the Boltzmann equation in the presence of gravitational fields. The Marle model \cite{Mar1,Mar2} of the collision operator of the Boltzmann equation is used and the transport coefficients of shear and bulk viscosities and thermal conductivity are determined by the method of Chapman-Enskog. It is shown that the transport coefficients in the presence of a gravitational field are smaller in comparison of their values in the absence of it.  Furthermore, the heat flux obtained has two relativistic terms,  aside from the temperature gradient. One of them  -- caused by the inertia of energy -- represents an isothermal heat flux when matter is accelerated and was suggested by Eckart \cite{Eck}. The other is proportional to the gravitational potential gradient and was proposed by Tolman \cite{To1,To2}. It expresses  that a state of equilibrium of a relativistic gas in a gravitational field and in the absence of an acceleration field  can be  attained only if the temperature gradient is counterbalanced by a gravitational potential gradient.

The work is structured as follows. In Section 2 the Marle model of the Boltzmann equation, the equilibrium distribution function in the Schwarzschild metric are introduced as well as the definitions and balance equations for the particle four-flow and energy-momentum tensor. The non-equilibrium distribution function, obtained by the use of Chapman-Enskog method, is the subject of Section 3.  In this section the Euler equations of a relativistic gas in a gravitational field are obtained and the constitutive equations with the corresponding transport coefficients are determined for a viscous heat-conducting relativistic gas. In Section 4 we state the main conclusions of this work and in the appendices we introduce the isotropic Schwarzschild metric and show how to evaluate the integrals when the distribution function depends on the components of the metric tensor.

 \section{Boltzmann equation}
Let us consider a single non-degenerate gas in a Riemannian space with line element $ds^2=g_{\mu\nu} dx^\mu dx^\nu$  where $g_{\mu\nu}$ is the metric tensor. A particle with rest mass $m$ is characterized by the space-time coordinates $(x^\mu)=(x^0=ct, {\bf x})$ and  by the momentum four-vector $(p^\mu)=(p^0, {\bf p})$, where $c$ denotes the speed of light.

The constant length of the momentum four-vector
$g_{\mu\nu}p^\mu p^\nu=m^2c^2$ implies that $p^0=\frac{p_0-g_{0i}p^i} {g_{00}}$, where
\ben\lb{1}
p_0=\sqrt{g_{00}m^2c^2+\left(g_{0i}g_{0j}-g_{00}g_{ij}\right)p^ip^j}.
\een
The components of the four-velocity with $U^\mu U_\mu=c^2$ are
 \be
 U^\mu=\left(\Gamma c,\Gamma v^i\right), \quad \Gamma=\frac{1}{\sqrt{g_{00}\left(1+\frac{g_{0i}}{g_{00}}\frac{v^i}{c}\right)^2-\frac{v^2}{c^2}}},
 \ee{2}
and in a comoving frame $\bf v=0$ yields $U^\mu=\left(\frac{c}{\sqrt{g_{00}}},\bf{0}\right)$.

 For the isotropic Schwarzschild metric (\ref{a2}) we have
 \ben\lb{3a}
 p^0=\frac{p_0}{g_0}\qquad p_0=\sqrt{g_0}\sqrt{m^2c^2+g_1\vert\bp\vert^2}, \\\lb{3b}
  U^\mu=\left(\frac{c}{\sqrt{g_0}},\bf{0}\right),\qquad \sqrt{-g}=\sqrt{g_0g_1^3},
 \een
where $g=\det(g_{\mu\nu})$.

In the phase space spanned by the space-time and momentum coordinates, the state of a relativistic gas is characterized by the one particle distribution function
$f(x^\alpha, p^\alpha)=f({\bf x}, {\bf p}, t)$, such that   $f({\bf x, p},t)d^3{ x}\, d^3{ p}$
 gives the number of particles  in the volume element $d^3x$ about ${\bf x}$ and with momenta in a range $d^3p$ about $\bf p$ at time $t$.

 The space-time evolution of the one-particle distribution function is governed by the Boltzmann equation.  Here for simplicity we shall use a model equation for the Boltzmann equation, which replaces the  collision term $Q(f,f)$ of the Boltzmann equation by a collision model $J(f)$. Two important model equations for  relativistic gases were proposed by Marle \cite{Mar1,Mar2} and Anderson and Witting \cite{AW}. The Marle model equation is given by
 \ben\label{m1}
 J(f)=-{m\over \tau}(f-f^{(0)}),
 \een
 while the Anderson-Witting model equation reads
 \ben\label{m2}
J(f)=-{U^{\mu}_{ L}p_{\mu}\over c^2\tau}(f-f^{ (0)}).
\een
  In the above equations, $\tau$ denotes a mean free time and $f^{ (0)}$ the Maxwell-J\"uttner distribution function. In the Marle model   the Eckart decomposition for the particle four-flow and energy-momentum tensor is used, while in the Anderson and Witting model the Landau-Lifshitz decomposition for these fields are employed. Note that in (\ref{m2}) the four-velocity $U_L$ refers to the Landau-Lifshitz decomposition.

The collision terms of the model equations must fulfill the same properties as those of the true collision term of the Boltzmann equation, namely
\begin{itemize}
\item[1)] For the summational invariant  $\psi=p^\mu$,
$Q(f,f)$ and $J(f)$ must satisfy the relationship
\ben\label{m3}
\int \psi Q(f,f)\sqrt{-g}{d^3p\over p_0}=0, \qquad\hbox{hence}\qquad
\int \psi J(f)\sqrt{-g}{d^3p\over p_0}=0;
\een
\item[2)] The  $\cal H$-theorem, or equivalently the tendency of the one-particle distribution function to the
equilibrium distribution function,  must hold
\ben\label{m4}
\int  Q(f,f)\ln f \sqrt{-g}{d^3p\over p_0}\leq 0, \qquad\hbox{hence}\qquad
\int J(f)\ln f\sqrt{-g} {d^3p\over p_0}\leq 0.
\een
\end{itemize}
For the proof of these properties one is referred e.g. to the book \cite{CK}.

  Here  we shall use the Marle model of the Boltzmann equation, which in the presence of gravitational fields reads
\be
p^\mu\frac{\partial f}{\partial x^\mu}-\Gamma_{\mu\nu}^ip^\mu p^\nu
\frac{\partial f}{\partial p^i}=-\frac{m}{\tau}\left(f-f^{ (0)}\right).
\ee{4}
Above, $\Gamma_{\mu\nu}^i$ are the Christoffel symbols and in a comoving frame the Maxwell-J\"uttner distribution function becomes
\be
f^{(0)}=\frac{n}{4\pi k T m^2 c  K_2(\z)}\exp\left({-\frac{c\sqrt{m^2c^2+g_1\vert\bp\vert^2}}{k T}}\right),
\ee{5}
thanks to (\ref{3a})$_2$ and (\ref{3b})$_1$.
Here $n, T$ and $k$ represent the particle number density, temperature and Boltzmann's constant, respectively. Furthermore, $K_2(\z)$ is the modified Bessel function of second kind
\be
 K_n(\zeta)=\left(\frac{\zeta}{2}\right)^n\frac{\Gamma(1/2)}{\Gamma(n+1/2)}\int_{1}^\infty e^{-\zeta y}\left(y^2-1\right)^{n-1/2}\,dy,
\ee{6}
which is a function of $\zeta=mc^2/kT$ representing the ratio of the particle rest energy $mc^2$  and the thermal energy of the gas $kT$. The limiting case $\z\gg1$ corresponds to a gas in the non-relativistic regime at low temperatures, while $\z\ll1$ refers to a ultra-relativistic gas at high temperatures.

The macroscopic state of a relativistic gas in a gravitational field may be described by the two first moments of the distribution function, the particle four-flow $N^\mu$ and the energy-momentum tensor $T^{\mu\nu}$. Their definitions in terms of the one-particle distribution function are given by
\be
N^\mu= c\int p^\mu f\,\sqrt{-g}\,\frac{d^3p}{p_0},\quad
T^{\mu\nu}= c\int p^\mu p^\nu f\,\sqrt{-g}\,\frac{d^3p}{p_0}.
\ee{7}
The respective balance equations are obtained from the Boltzmann equation, yielding
\be
{N^\mu}_{;\mu}= 0,\qquad  {T^{\mu\nu}}_{;\mu}=0,
\ee{8}
where the semicolon denotes a covariant derivative.

It is usual in the thermodynamic theory of relativistic fluids to decompose the  particle four-flow and the energy-momentum tensor in terms of quantities that appear in the theory of non-relativistic fluid dynamics, namely,  particle number density $n$, energy per particle $e$, hydrostatic pressure $\p$, non-equilibrium pressure $\varpi$, pressure deviator $\mathbb{P}^{\mu\nu}$ (the traceless part of the pressure tensor) and heat flux $q^\mu$. For Marle's model equation the decomposition used is that of Eckart (see e.g. \cite{Eck,CK})
\ben\lb{9a}
N^{\mu}&=&nU^{\mu},\\\lb{9b}
T^{\mu\nu}&=&\mathbb{P}^{\mu\nu}-\left(\p+\varpi\right)
\Delta^{\mu\nu}
+\frac{1}{c^2}\left(q^\mu U^\nu+q^\nu U^\mu\right)+\frac{en}{c^2}U^{\mu} U^{\nu},
\een
where $\Delta^{\mu\nu}$ is the projector
\be
\Delta^{\mu\nu}=g^{\mu\nu}-\frac{1}{c^2}U^\mu U^\nu.
\ee{10}

From the insertion of the Maxwell-J\"uttner distribution function (\ref{5}) into the definition of the energy-momentum tensor (\ref{7})$_2$ and integration of the resulting equation it follows that the energy per particle and the hydrostatic pressure read
\be
e=mc^2\left[\frac{K_3}{K_2}-\frac{1}{\zeta}\right],\qquad \p=nkT,
\ee{10a}
respectively.

\section{Chapman-Enskog method}

In the Chapman-Enskog method the distribution function is written as $f=f^{(0)}\left(1+\varphi\right)$, where $\varphi$ -- the deviation from the Maxwell-J\"uttner distribution function -- is considered as a small quantity, i.e., $\vert\varphi\vert<1$. Furthermore, the  Maxwell-J\"uttner distribution function is inserted on the left-hand side of the Boltzmann equation (\ref{4}) and the representation $f=f^{(0)}\left(1+\varphi\right)$ on its right-hand side. By performing the derivatives it follows
\ben\no
&&-\frac{m}{\tau}\left(f-f^{ (0)}\right)=f^{ (0)}\bigg\{\frac{p^\nu}{n}\frac{\partial n}{\partial x^\nu}
+\frac{p^\nu}{T}\left[1-\frac{K_3\z}{K_2}+\frac{ p^\tau U_\tau}{kT}\right]\frac{\partial T}{\partial x^\nu}
\\\lb{11}
&&-\frac{p^i p^\nu}{kT}\frac{\partial U_i}{\partial x^\nu}
-\frac{c^2}{2kT}\frac{d g_1}{dr}\frac{p^ip^jp^k}{U^\tau p_\tau}\delta_{ij}\delta_{kl}\frac{x^l}{r}
+\frac{c^2}{k T}g_1\delta_{ij}\Gamma_{\sigma\nu}^i \frac{p^j p^\sigma p^\nu }{U^\tau p_\tau}\bigg\}=-\frac{m}{\tau}f^{ (0)}\varphi.
\een
Hence, the deviation from the Maxwell-J\"uttner distribution $\varphi$ is determined as a function of gradients and  derivatives of the metric tensor components.

\subsection{Euler's equations}

A relativistic fluid in the absence of the gradients of temperature and four-velocity represents an Eulerian  fluid. The determination of Euler's equations proceed as follows.
First the multiplication of (\ref{11}) by $\sqrt{-g}d^3 p/p_0$ and integration of the resulting equation leads to the balance equation of particle number density, namely,
\be
U^\nu\,\frac{\partial n}{\partial x^\nu}+n\, {U^\nu}_{;\nu}=0.
\ee{12a}
Next the multiplication of (\ref{11}) by $p^\mu\sqrt{-g}d^3 p/p_0$ and subsequent integration  implies into an equation which is used to derive the balance equations for the energy density and momentum density of an Eulerian fluid. The energy density balance equation is obtained through the projection $U_\mu$, yielding
\ben\label{12b}
n \,c_v\,U^\nu\,\frac{\partial T}{\partial x^\nu}+\p\, {U^\nu}_{;\nu}=0,
\een
where $c_v$ is the heat capacity per particle at constant volume
\ben\lb{12c}
c_v=\frac{\partial e}{\partial T}=k\left(\zeta^2+5\frac{K_3}{K_2}\zeta-\frac{K_3^2}{K_2^2}\zeta^2-1\right).
\een
The momentum density balance equation results from the projection $\Delta^\nu_\mu$:
\ben\lb{12d}
m\,n\,\frac{K_3}{K_2}U^\mu \frac{\partial U_i}{\partial x^\mu}-\frac{\partial \p}{\partial x^i}
-m\,n\,\frac{K_3}{K_2}\frac{1}{1-\Phi^2/4c^4}\frac{\partial\Phi}{\partial x^i}=0,
\een
where we have introduced the gravitational potential
\ben\lb{13}
\Phi=-\frac{GM}{r},\quad\hbox{with} \quad \frac{\partial \Phi}{\partial x^k}=\frac{GM}{r^2}\delta_{kj}\frac{x^j}{r}.
\een

We note that (\ref{12d}) is a function of the ratio $\vert\Phi(r)\vert/c^2$, which we can  estimate at the surface of some bodies:
\begin{enumerate}
\item  Earth: $M_{\oplus}\approx 5.97\times
10^{24}$ kg; $R_\oplus\approx 6.38\times 10^6$ m;
${\vert\Phi(R_\oplus)\vert/c^2}\approx 7 \times 10^{-10}$;
\item   Sun: $M_{\odot}\approx 1.99\times
10^{30}$ kg; $R_\odot\approx 6.96\times 10^8$ m; ${\vert\Phi(R_\odot)\vert/c^2}\approx 2.2 \times 10^{-6}$;
\item   White dwarf: $M\approx 1.02 M_{\odot}$;
$R\approx 5.4\times 10^6$ m;
${\vert\Phi(R)\vert/c^2}\approx 2.8 \times 10^{-4}$;
\item   Neutron  star: $M\approx
M_{\odot}$; $R\approx 2\times 10^4$ m;
${\vert\Phi(R)\vert/c^2}\approx 7.5 \times 10^{-2}$.
\end{enumerate}
The above estimates imply that in most cases the approximation $\vert\Phi\vert/c^2\ll1$ is valid. Hence, the momentum density balance equation (\ref{12d}) in the non-relativistic limiting case $\zeta\gg1$ and in the presence of a weak gravitational field reads
\ben\label{14}
m\,n\,\left(1+\underline{\frac{5}{2\z}+\dots}\right)U^\mu \frac{\partial U_i}{\partial x^\mu}-\frac{\partial \p}{\partial x^i}
-m\,n\,\left(1+\underline{\frac{5}{2\z}+\dots}\right)\left(1+\underline{\frac{\Phi^2}{4c^4}+\dots}\right)\frac{\partial\Phi}{\partial x^i}=0.
\een
Note that without the underlined terms (\ref{14}) reduces to the usual form of Newton's second law for a non-relativistic gas  in the presence of a weak gravitational field.

\subsection{Viscous and heat-conducting relativistic gas}

In order to determine the energy-momentum  tensor  with the deviation of the  Maxwell-J\"uttner distribution function $\varphi$ -- given by (\ref{11}) -- we
first note that according to (\ref{m1}) and (\ref{m3}) the particle four-flow evaluated with the equilibrium and non-equilibrium distribution functions must have the same representation, i.e., $N^\mu=N_E^\mu=nU^\mu$. From now on the index $E$ will denote the value of a quantity at equilibrium, i.e., when  evaluated with the Maxwell-J\"uttner distribution function. However, in Marle's model the pressure and the energy per particle have different values at equilibrium and non-equilibrium states. This is not the case of the  Anderson and Witting model equation (\ref{m2}) and of the full Boltzmann equation (see e.g. \cite{CK}).

In the following we shall write the constitutive equations in a comoving frame where the components of the projector become
\be
\Delta^{00}=0,\qquad \Delta^{ij}=g^{ij}=-\frac{1}{g_1}\delta^{ij}=-\frac{1}{\left(1+\frac{\vert\Phi\vert}{2c^2}\right)^4}\delta^{ij}.
\ee{18}

The knowledge of the  energy-momentum tensor  is obtained from  the insertion of the representation of the distribution function $f=f^{(0)}(1+\varphi)$ -- with $\varphi$ given by (\ref{11}) -- into its definition (\ref{7})$_2$ and subsequent integration of the resulting equation.

Let us first analyze  the projections
\ben\lb{16}
\varpi+p=-\frac{1}{3}\Delta_{\mu\nu}\,T^{\mu\nu},\qquad ne=\frac{1}{c^2}U_\mu U_\nu T^{\mu\nu},
\een
which gives the following relationships
\ben\label{16a}
\varpi+(\p-\p_E)=\frac{\tau k\p_E}{3c_v^E\left(1+\frac{\vert\Phi\vert}{2c^2}\right)^4}\left[20\frac{K_3^E}{K_2^E}+3\zeta_E+2\zeta_E^2\left(\frac{K_3^E}{K_2^E}\right)^3
%\right.\\\left.
-13\zeta_E
\left(\frac{K_3^E}{K_2^E}\right)^2-2\zeta_E^2\frac{K_3^E}{K_2^E}\right]\,\frac{\partial U^j}{\partial x^j},
\\\label{16b}
(e-e_E)=\frac{\tau mc^2k}{c_v^E\zeta_E\left(1+\frac{\vert\Phi\vert}{2c^2}\right)^4}\left[20\frac{K_3^E}{K_2^E}+3\zeta_E+2\zeta_E^2\left(\frac{K_3^E}{K_2^E}\right)^3
%\right.\\\left.
-13\zeta_E
\left(\frac{K_3^E}{K_2^E}\right)^2-2\zeta_E^2\frac{K_3^E}{K_2^E}\right]\,\frac{\partial U^j}{\partial x^j}.
\een
In the above equations it was  introduced  $K_n^E\equiv K_n(\zeta_E)$.

We can relate the  pressure difference $(p-p_E)$ with the energy difference $(e-e_E)$ by noting that pressure difference can be written  as a temperature difference through  $(p-p_E)=nk(T-T_E)=nmc^2(1/\zeta-1/\zeta_E)$.  If we expand  the energy difference $(e-e_E)$ about $T_E$ and neglect terms up to second order in $(T-T_E)$ -- since we are only interested in processes close to equilibrium with gradients of first order -- it follows that
$e-e_E=c_v^E(T-T_E)$. Hence these two last relationship together with (\ref{16}) furnishes:
\ben\label{16c}
p-p_E=\frac{\tau\p_E k^2}{(c_v^E)^2\left(1+\frac{\vert\Phi\vert}{2c^2}\right)^4}\left[20\frac{K_3^E}{K_2^E}+3\zeta_E+2\zeta_E^2\left(\frac{K_3^E}{K_2^E}\right)^3
%\right.\\\left.
-13\zeta_E
\left(\frac{K_3^E}{K_2^E}\right)^2-2\zeta_E^2\frac{K_3^E}{K_2^E}\right]\,\frac{\partial U^j}{\partial x^j}.
\een

Now we can obtain from (\ref{16a}) and (\ref{16c}) the final expression for the non-equilibrium pressure
\ben\label{19}
\varpi=-\eta\,\frac{\partial U^j}{\partial x^j},
\een
where $\eta$ denotes the coefficient of bulk viscosity, which is given by
 \ben\nonumber
&&\eta=\frac{\tau k^2\p_E}{3(c_v^E)^2 \left(1+\frac{\vert\Phi\vert}{2c^2}\right)^4}\left[20\frac{K_3^E}{K_2^E}-13\left(\frac{K_3^E}{K_2^E}\right)^2\z_E-2\frac{K_3^E}{K_2^E}\z_E^2
+3\z_E\right.
\\\lb{22a}
&&\left.+2\left(\frac{K_3^E}{K_2^E}\right)^3\z_E^2\right]\left(4-\z_E^2-5\frac{K_3^E}{K_2^E}\z_E+\left(\frac{K_3^E}{K_2^E}\right)^2\z_E^2\right).
\een

The heat flux and the traceless part of the pressure tensor are obtained from the projections
\ben\lb{17}
&&q^\sigma=\Delta_\mu^\sigma\, U_\nu\,T^{\mu\nu},\qquad\mathbb{P}^{\sigma\tau}=\left[\Delta^{\sigma}_{\mu}\Delta^{\tau}_{\nu}-\frac{1}{3}\Delta^{\sigma\tau}\Delta_{\mu\nu}
\right]T^{\mu\nu},
\een
and it it follows the constitutive equations
\ben\label{21}
&&q^i=-\lambda \delta^{ij}\left[\frac{\partial T}{\partial x^j}-\frac{T}{c^2}U^\sigma \frac{\partial U_j}{\partial x^\sigma}+\frac{T}{c^2}\frac{1}{1-\Phi^2/4c^4}\frac{\partial\Phi}{\partial x^j}\right],
\\\lb{20}
&&\mathbb{P}^{ij}=-\mu \Bigg[\left(\delta^{ik}\delta^{jl}
+\delta^{il}\delta^{jk}\right)
-\frac{2}{3} \delta^{ij}\delta^{kl}
\Bigg]\frac{\partial U_k}{\partial x^l}.
\een
The coefficients $\lambda$ and $\mu$ are identified with the thermal conductivity and shear viscosity, respectively, and their expressions read
\ben\lb{23}
&&\lambda=\frac{\tau \p_E}{\left(1+\frac{\vert\Phi\vert}{2c^2}\right)^4} \frac{k}{m}\z_E\left(\z+5\frac{K_3^E}{K_2^E}-\left(\frac{K_3^E}{K_2^E}\right)^2\z_E\right),
\\\lb{23a}
&&\mu=\frac{\tau\p_E}{\left(1+\frac{\vert\Phi\vert}{2c^2}\right)^8}\frac{K_3^E}{K_2^E}.
\een

Equations (\ref{19}) and (\ref{20}) represent the constitutive equations of a Newtonian relativistic fluid, while (\ref{21}) the generalized Fourier law.

Let us analyze the transport coefficients given by (\ref{22a}), (\ref{23}) and (\ref{23a}). All these  coefficients depend on gravitational potential through the component  of the metric tensor $g_1(r)=\left(1+\frac{GM}{2c^2r}\right)^4$ and we may infer from their expressions that the values become smaller in presence of a gravitational field. The decrease of the transport coefficients due to the gravitational field is rather small for stellar objects which are not too compact, i.e., for stellar objects where $\vert\Phi\vert/c^2\ll1$. In the absence of the gravitational potential the metric tensor reduces to the  one of a Minkowski space-time. In this case $g_1=1$ and the expressions for the transport coefficients reduce to those of   Marle \cite{Mar2,CK}.

In the limiting case of weak gravitational fields $\vert\Phi\vert/c^2\ll1$ and low temperatures $\z_E\gg1$ the transport coefficients (\ref{22a}), (\ref{23}) and (\ref{23a}) become
\ben
\eta=\frac{5\p_E\tau}{6\z^2}\left[1-\frac{21}{2\z_E}+\dots\right]\left[1-\frac{2\vert\Phi\vert}{c^2}+\dots \right],\\
\mu=\p_E\tau\left[1+\frac{5}{2\z_E}+\dots\right]\left[1-\frac{4\vert\Phi\vert}{c^2}+\dots \right],\\
\lambda=\frac{5k\p_E\tau}{2m}\left[1+\frac{3}{2\z_E}+\dots\right]\left[1-\frac{2\vert\Phi\vert}{c^2} +\dots\right],
\een
The above expressions  correspond to a non-relativistic gas in a weak gravitational field.

The transport coefficients (\ref{22a}), (\ref{23}) and (\ref{23a}) in the limiting case of weak gravitational fields $\vert\Phi\vert/c^2\ll1$ and high temperatures $\z_E\ll1$ read
\ben
&&\eta=\frac{\p_E\tau\z_E^3}{54}\left[1+\left(\frac{31}{12}+\frac{9}{2}\ln\left(\frac{\z_E}{2}\right)+\frac{9}{2}\gamma\right)\zeta_E^2+\dots\right]
\left[1-\frac{2\vert\Phi\vert}{c^2}+\dots \right],\qquad\\
&&\mu=\frac{4\p_E\tau}{\z_E}\left[1+\frac{\z_E^2}{8}+\dots\right]\left[1-\frac{4\vert\Phi\vert}{c^2}+\dots \right],\\
&&\lambda=\frac{4c^2\p_E\tau}{T\z_E}\left[1-\frac{\z_E^2}{8}+\dots\right]\left[1-\frac{2\vert\Phi\vert}{c^2} +\dots\right].
\een
Here the expressions  refer to a gas in a weak gravitational field and in the ultra-relativistic regime.

As a  remark we call attention to the fact  that in the literature of neutrons stars (see e.g. \cite{CHa})
the dependence of the transport coefficients on the gravitational potential is not taken into account.

\subsection{Fourier's law }

Let us analyze Fourier's law with more details. It has three terms, namely,
 \begin{description}
   \item[1] the usual dependence of the heat flux in the gradient of temperature $\frac{\partial T}{\partial x^j}$;
   \item[2] a relativistic term proportional to $\frac{T}{c^2}U^\sigma \frac{\partial U_j}{\partial x^\sigma}$ which represents an isothermal heat flux when matter is accelerated. It is a  small term
   that acts in a direction opposite to the acceleration and is due to the inertia of energy. It was proposed by Eckart \cite{Eck} within an irreversible thermodynamic theory;
   \item[3] the third term $\frac{T}{c^2}\frac{1}{1-\Phi^2/4c^4}\frac{\partial\Phi}{\partial x^j}$ is proportional to the gravitational potential gradient. It was proposed by Tolman \cite{To2} and indicates that in the absence of the  acceleration term  a state of equilibrium of a relativistic gas in a gravitational field can be attained if the temperature gradient is counterbalanced by a gravitational potential gradient. In a weak gravitational field  the equilibrium condition reads
       \be
       \frac{1}{T}\nabla T=-\frac{\bf g}{c^2},
       \ee{24}
       which is the so-called Tolman's law with $\bf g$ being the gravitational field.
    \end{description}

    By considering that the temperature field depends only on the radial coordinate $T=T(r)$ and that there is no heat flux and acceleration field, it follows from (\ref{21}) that the temperature field in a gravitational field must fulfill the differential equation
       \be
       \frac{1}{T(r)}\frac{dT(r)}{dr}=-\frac{1}{c^2}\frac{1}{1-\Phi(r)^2/4c^4}\frac{d\Phi(r)}{dr}.
       \ee{25}
       The solution of the above differential equation for the boundary condition $T(R)=1$ -- where $R$ is the radius of the spherical source -- is
       \be
       T(r)=\frac{\left(1-\frac{\vert\Phi(R)\vert}{2c^2}\right)\left(1+\frac{\vert\Phi(R)\vert}{2c^2}\frac{R}{r}\right)}{\left(1+\frac{\vert\Phi(R)\vert}{2c^2}\right)
       \left(1-\frac{\vert\Phi(R)\vert}{2c^2}\frac{R}{r}\right)},
       \ee{26}
       which is a  decreasing function of $r$ for  $r>R$. In a weak gravitational field where $\vert\Phi(R)\vert/c^2$ is a small quantity (\ref{26}) can be expressed as
       \be
       T(r)=1-\frac{\vert\Phi(R)\vert}{c^2}\left(1-\frac{R}{r}\right)+\frac{\vert\Phi(R)^2\vert}{2c^2}\left(1-\frac{R}{r}\right)^2+\dots\,.
       \ee{27}

    It is worth to call attention that if we use the momentum density balance equation of an Eulerian fluid (\ref{12d}) in order to eliminate the acceleration term we get that the heat flux (\ref{21}) becomes
    \be
    q^i=-\lambda \delta^{ij}\left[\frac{\partial T}{\partial x^j}-\frac{TK_2}{nmc^2K_3}\frac{\partial \p}{\partial x^j}\right].
    \ee{28}
    We can transform the dependence of the pressure gradient in (\ref{28}) into a dependence on the temperature gradient and particle number density gradient through the use of the equation of state $\p=nkT$  and in this new description the presence of the gravitational potential gradient disappears. This indicates that the term related with the gravitational potential gradient  in the heat flux is associated with the acceleration term.   However, in the work \cite{AL}  the acceleration term was eliminated by the use of the momentum balance equation and Fourier's law  has a dependence on the gravitational field and of the gradient of particle number density. We presume that the difference of the results  is due to the fact that in the referred work there is no dependence of the equilibrium distribution function on the metric tensor components.

\section{Conclusions}

In this work we have analyzed a relativistic gas in a Schwarzschild metric within the framework of the Boltzmann equation in the presence of  gravitational fields. The model of Marle for the collision operator of the Boltzmann equation was used and the deviation from the equilibrium distribution function was determined from the Chapman-Enskog method. Among other results it was shown: \textbf{(i)} apart from the temperature gradient the heat flux has two relativistic terms, one due to the inertia of the energy is connected with an acceleration, while the other is proportional to the gravitational potential gradient; \textbf{(ii)} the transport coefficients depend on the gravitational potential and become smaller than their values in the absence of it; \textbf{(iii)} expressions for the transport coefficients in the limiting cases of low and high temperatures for weak gravitational fields were given; \textbf{(iv)} the radial temperature field of the gas is a decreasing function of the distance of the source of the gravitational field, when the heat flux and acceleration field vanish; \textbf{(v)} the momentum balance equation for an Eulerian fluid in the presence of a weak gravitational field and in the non-relativistic regime reduces to the usual expression of Newton's second law.

As a final comment it is worth to call attention that the present theory should not be valid in the case where   $\Phi\rightarrow2c^2$. In this case the the transport coefficients diverge and we are at the presence of the horizon of a Schwarzschild black hole. This limiting case will be the subject of a future  investigation.

\appendix
\section{Appendix: Schwarzschild metric}
The  Schwarzschild metric is the solution of Einstein's field equation for a spherical symetrical non-rotating and uncharged source of the gravitational field with total mass $M$. In terms of the spherical coordinates $(\breve r, \theta, \varphi)$ it is given by
 \ben\label{a1}
 ds^2={\left(1-\frac{2GM}{c^2\breve r}\right)}\left(dx^0\right)^2-\frac{1}{\left(1-\frac{2GM}{c^2\breve r}\right)}d\breve r^2
  -\breve r^2\left(d\theta^2+\sin^2\theta d\phi^2\right).
 \een
 where $G$ is the gravitational constant. The isotropic Schwarzschild metric (see e.g. \cite{Bu}) follows by introducing a new radial coordinate $r^2=\delta_{ij}x^ix^j$ through the relationship
 $\breve r=r\left(1+\frac{GM}{2c^2r}\right)^2$  so that (\ref{a1}) becomes
 \ben\lb{a2}
 ds^2=g_0(r)\left(dx^0\right)^2-g_1(r)\delta_{ij}dx^idx^j.
 \een
 Above we have introduced the abbreviations
 \be
g_0(r)=\frac{\left(1-\frac{GM}{2c^2r}\right)^2}{\left(1+\frac{GM}{2c^2r}\right)^2},\qquad
g_1(r)=\left(1+\frac{GM}{2c^2r}\right)^4.
\ee{a3}

For the isotropic metric (\ref{a2}) the Cristoffel symbols read
 \ben\lb{a4a}
&& \Gamma_{00}^0=0,\qquad \Gamma_{ij}^0=0,\qquad \Gamma_{ij}^k=0 \quad (i\neq j\neq k),\qquad
\Gamma_{0j}^i=0,
 \\\lb{a4b}
  &&\Gamma_{\underline{i}\,j}^{\underline{i}}=\frac{1}{2g_1(r)}\frac{d g_1(r)}{dr}\delta_{jk} \frac{x^k}{r},
 \qquad\Gamma_{0i}^0=\frac{1}{2g_0(r)}\frac{d g_0(r)}{dr}\delta_{ij}
 \frac{x^j}{r},
 \\\lb{a4d}
 &&\Gamma_{00}^i=\frac{1}{2g_1(r)}\frac{d g_0(r)}{dr}\frac{x^i}{r},\qquad
 \Gamma_{\underline{i}\,\underline{i}}^j=-\frac{1}{2g_1(r)}\frac{d g_1(r)}{dr}\frac{x^j}{r}\quad (i\neq j),
 \een
where the underlined indices are not summed and
\ben\lb{a5a}
\frac{d g_0(r)}{dr}=\frac{2GM}{c^2r^2}\frac{\left(1-\frac{GM}{2c^2r}\right)}{\left(1+\frac{GM}{2c^2r}\right)^3},\qquad
\frac{d g_1(r)}{dr}=-\frac{2GM}{c^2r^2}\left(1+\frac{GM}{2c^2r}\right)^3.
\een

\section{Appendix: Evaluation of integrals}

 To evaluate the following integral
\be
Z=\sqrt{-g}\int e^{-\frac{1}{kT}U^\lambda p_\lambda}\frac{d^3p}{p_0},
\ee{a6}
we choose a comoving frame so that the above integral reduces to
\be
Z=4\pi\sqrt{g_0g_1^3}\int_0^\infty e^{-{c p_0}/(kT\sqrt{g_0})}\vert {\bf p}\vert^2
\frac{d\vert {\bf p}\vert}{p_0},
\ee{a7}
through the introduction of spherical coordinates $d^3p=\vert {\bf p}\vert^2\sin\theta
d\vert {\bf p}\vert d\theta d\varphi$ and by integrating in the angles
 $0\leq\theta\leq\pi$ and $0\leq\varphi\leq 2\pi$. If we use the
relationship (\ref{3a})$_2$ and change the variable
of integration through $p_0=mc\,y\sqrt{g_0}$ we have that
\be
\vert {\bf p}\vert^2=\frac{m^2c^2}{g_1}(y^2-1),\qquad
d\vert {\bf p}\vert=\frac{mc}{\sqrt{g_1}}\frac{y\,dy}{(y^2-1)^\frac{1}{2}},
\ee{a8}
and (\ref{a7}) reduces to
\be
Z=4\pi m^2c^2\int_1^\infty e^{-\zeta y}(y^2-1)^\frac{1}{2}
dy =4\pi m kT
K_1(\zeta),
\ee{a9}
i.e., $Z$ is given in terms of the modified Bessel function of second kind.

For the evaluation of the integral
\be
Z_1=\sqrt{-g}\int \frac{e^{-\frac{1}{kT}U^\lambda p_\lambda}}{U^\tau p_\tau}\frac{d^3p}{p_0},
\ee{a10}
we change the variables and introduce
\be
p_0=mc\sqrt{g_0}\cosh t,\qquad \vert{\bf p}\vert=\frac{mc}{\sqrt{g_1}}\sinh t,
\ee{a11}
and get
\ben\lb{a12}
Z_1=4\pi m\int_0^\infty \frac{\cosh^2t-1}{\cosh t}e^{-\z\cosh t}\,dt
=4\pi m\left[K_1(\zeta)-{\rm Ki}_1(\zeta)\right],
\een
thanks to the definition
\be
{\rm Ki}_n(\zeta)=\int_\zeta^\infty {\rm Ki}_{n-1}(t)dt=
\int_0^\infty \frac{e^{-\zeta\cosh t}}{\cosh^nt}dt.
\ee{a13}

Following the same methodology it is easy to obtain
\be
Z^{\mu}=\sqrt{-g}\int p^{\mu}e^{-\frac{1}{kT}U_\lambda p^\lambda}\frac{d^3p}{p_0}=
4\pi m^2k T K_2(\zeta)U^{\mu},
\ee{a14}
\be
Z_1^\mu=\sqrt{-g}\int p^\mu\frac{e^{-\frac{1}{kT}U^\lambda p_\lambda}}{U^\tau p_\tau}\frac{d^3p}{p_0}=4\pi m^2\frac{K_1(\zeta)}{\z}U^\mu,
\ee{a15}
and to derive further expressions for $Z^{\mu_1\dots\mu_n}$ and $Z_1^{\mu_1\dots\mu_n}$.

\section*{Acknowledgments}
 This paper was partially supported by Conselho Nacional de Desenvolvimento Cient\'{\i}fico e Tecnol\'ogico (CNPq),  Brazil.

%%%%%%%%%%%%%%%%%%%%%%%%%%%%%%%%%%%%%%%%%%%%%%%%%%%%%%%%%%%%%%%%%%%%%%%%%%


\begin{thebibliography}{99}
%%%%%%%%%%%%%%%%%%%%%%%%%%%%%%%%%%%%%%%%%%%%%%%%%%%%%%%%%%%%%%%%%%%%%%%%%%

\bibitem{J1}
 J\"uttner F, \emph{Das Maxwellsche Gesetz der Geschwindigkeitsverteilung
in der Relativtheorie}, 1911 { Ann. Physik und Chemie} {\bf 34} 856

\bibitem{J2}
 J\"uttner F, \emph{Die relativistische Quantentheorie des idealen Gases}, 1928
{Zeitschr. Physik} {\bf 47}, 542

\bibitem{LM}  Lichnerowicz A and  Marrot R, \emph{Propri\'et\'es statistiques des ensembles de particules en relativit\'e restreinte}, 1940  {C. R. Acad. Sci.}  \textbf{210} 759

\bibitem{I}   Israel W, \emph{Relativistic kinetic theory of a simple gas,} 1963 J. Math. Phys. {\bf4} 1163

\bibitem{K} Kelly  D C, 1963 {\it The Kinetic Theory of a Relativistic Gas},
unpublish report (Oxford: Miami University)

\bibitem{Ch1}  Chernikov N A, \emph{The relativistic gas in the gravitational field}, 1963 Acta Phys. Pol. {\bf23} 629

 \bibitem{Ch2}  Chernikov N A, \emph{Equilibrium distribution of the relativistic
gas}, 1964 Acta Phys. Pol. {\bf26} 1069

\bibitem{B}  Bernstein J, 1988 \emph{Kinetic Theory in the Expanding Universe},  (Cambridge:  Cambridge University Press).

\bibitem{And1}   Doi T,  Santos A and  Tij M, \emph{Numerical study of the influence of gravity on the heat conductivity on the basis of kinetic theory}
 1999 Phys. Fluids {\bf11} 3553

\bibitem{And2} Tij  M,  Garz\'o V and  Santos A, \emph{On the influence of gravity on the thermal
conductivity},  in \emph{Rarefied Gas Dynamics},  Brun R,  Campargue R,  Gatignol R,
and  Lengrand J-C, eds. 1999 (Toulouse: C\'epadu\`es) p 239

\bibitem{To1} Tolman  R C, \emph{On the weight of heat and thermal equilibrium
in general relativity}, 1930 Phys. Rev. {\bf35} 904

\bibitem{To2} Tolman  R C, \emph{Temperature equilibrium in a static
gravitational field}, 1930 Phys. Rev. {\bf36} 1791

\bibitem{AL}  Sandoval-Villalbazo A,  Garcia-Perciante A L and  Brun-Battistini D, \emph{Tolman's law in linear irreversible thermodynamics: a kinetic theory approach}, 2012 Phys. Rev. D {\bf86} 084015

\bibitem{Mar1}  Marle C, \emph{Sur l'\'etablissement des \'equations de l'
hydrodynamique des fluides relativistes dissipatifs, I. L'\'equation de
Boltzmann relativiste}, 1969  { Ann. Inst. Henri Poincar\'e}
{\bf 10} 67

\bibitem{Mar2}  Marle C, \emph{Sur l'\'etablissement des \'equations de l'
hydrodynamique des fluides relativistes dissipatifs, II. M\'ethodes de
r\'esolution approch\'ee de
l'\'equation de Boltzmann relativiste}, 1969 { Ann. Inst. Henri Poincar\'e}
{\bf 10} 127

\bibitem{Eck}  Eckart C, \emph{The thermodynamics of irreversible processes, III.
Relativistic theory of a simple fluid}, 1940 Phys. Rev. {\bf58}, 919

\bibitem{AW} J. L. Anderson and  H. R. Witting, \emph{A relativistic relaxation-time
model for the Boltzmann equation}, 1974 { Physica} {\bf 74}, 466


\bibitem{CK}   Cercignani C and  Kremer G M, 2002 {\it
The Relativistic Boltzmann Equation: Theory and Applications}
(Basel: Birkh\"auser)

\bibitem{CHa} N. Chamel and P. Haensel \emph{Physics of Neutron Star Crusts},
http://www.livingreviews.org/lrr-2008-10

\bibitem{Bu} Adler R,  Bazin M and  Schiffer M, 1965 \emph{Introduction to General Relativity} (New York: McGraw Hill)


\end{thebibliography}
\end{document}